\newif\ifpublic
\publictrue 

\ifpublic
\documentclass[11pt,letterpaper]{article}
\else
\documentclass[11pt,draft, letterpaper]{article}
\fi
\usepackage[margin=1in]{geometry}
\usepackage[final,pagebackref,colorlinks=true,linkcolor=blue,%
urlcolor=blue,citecolor=blue,pdfstartview=FitH]{hyperref}
\usepackage[T1]{fontenc}    
\usepackage{amsmath}       
\usepackage{amssymb}
\usepackage{amsthm}
\usepackage{empheq}
\usepackage{amsbsy}
\usepackage{nicefrac}  
\usepackage{braket}
\usepackage{lipsum}
\usepackage{fullpage}
\usepackage{mathpazo}
\usepackage{mathtools}
\usepackage{enumitem}
\usepackage{subfiles}
\usepackage{xcolor}
\usepackage{setspace}
\usepackage{fixltx2e}
\usepackage{algpseudocode}
\usepackage[linesnumbered,ruled, lined, commentsnumbered, longend]%
{algorithm2e}
\ifpublic
\usepackage[disable]{todonotes}
\else
\usepackage[colorinlistoftodos]{todonotes}
\fi

\usepackage[capitalize,nameinlink]{cleveref}
\crefname{algocf}{alg.}{algs.}
\Crefname{algocf}{Algorithm}{Algorithms}
\renewcommand{\eqref}[1]{\hyperref[#1]{(\ref*{#1})}}
\usepackage{stmaryrd}
\usepackage{MnSymbol} 
\usepackage{tikz}
\usepackage{pgfplots}
\usepackage{subcaption}
\usepackage{lipsum}
\usepackage{etoolbox}
\usepackage{appendix}
\usepackage{accents}
\usepackage{titlesec}
\titleclass{\subsubsubsection}{straight}[\subsection]
\newcounter{subsubsubsection}[subsubsection]
\renewcommand%
\thesubsubsubsection{\thesubsubsection.\arabic{subsubsubsection}}
 
\titleformat{\subsubsubsection}
  {\normalfont\normalsize\bfseries}{\thesubsubsubsection}{1em}{}
\titlespacing*{\subsubsubsection}
{0pt}{3.25ex plus 1ex minus .2ex}{1.5ex plus .2ex}
\AtBeginEnvironment{figure}{\addvspace{7mm}}%
\AtEndEnvironment{figure}{\addvspace{7mm}}
\pgfplotsset{compat=1.14}
\theoremstyle{plain}
\newtheorem{theorem}{Theorem}[section]

\newtheorem{lemma}[theorem]{Lemma}

\newtheorem{definition}[theorem]{Definition}

\newtheorem{fact}[theorem]{Fact}
\theoremstyle{definition}
\newtheorem{remark}[theorem]{Remark}

\newcounter{example}[section]

\crefname{subsubsubsection}{Section}{sections}

\renewcommand{\epsilon}{\varepsilon}

\newcommand{\sub}{\textsubscript}
\renewcommand{\phi}{\varphi}

\newcommand{\brak}[1]{\lbrace#1\rbrace}
\DeclareMathOperator{\E}{\mathbb{E}}

\newcommand{\norm}[1]{\left\lVert#1\right\rVert}

\newcommand{\imaxt}{\Tilde{I}_{\max}^{\epsilon}}

\DeclareMathOperator{\Tr}{Tr}
\DeclarePairedDelimiter{\abs}{\lvert}{\rvert}

\title{{\bf One-shot inner bounds for sending private classical information
over a quantum MAC}}

\author{
Sayantan Chakraborty${}^\dag$
\and
Aditya Nema\thanks{
Department of Mathematical Informatics,
Nagoya University, Japan.
Email: {\sf aditya.nema30@gmail.com}.
Supported by MEXT Quantum Leap Flagship Program (MEXT Q-LEAP), 
Grant Number JPMXS0120319794.
}
\and
Pranab Sen\thanks{
School of Technology and Computer Science, 
Tata Institute of Fundamental Research, Mumbai, India.
Email: {\sf
\{kingsbandz, pranab.sen.73\}@gmail.com
}
}
}

\date{}

\begin{document} 
\maketitle

\begin{abstract}
We provide the first inner bounds for sending private classical
information over a quantum multiple access channel.
We do so by using three powerful information theoretic
techniques: rate splitting, quantum simultaneous decoding for multiple
access channels, and a novel smoothed distributed covering
lemma for classical quantum channels. Our inner bounds are given in
the one shot setting and accordingly the three techniques used are
all very recent ones specifically designed to work in this setting.
The last technique is new to this work and is our
main technical advancement. For the asymptotic iid
setting, our one shot inner bounds
lead  to the natural quantum analogue of the best classical inner
bounds for this problem.
\end{abstract}

\section{Introduction}
Private communication over a noisy channel
is an important information processing and cryptographic primitive.
Here, a sender Alice wants to send her message over a noisy channel 
$\mathfrak{C}$ so that the genuine receiver Bob can decode it with
small error. At the same time, an eavesdropping receiver Eve should
get almost no information about the transmitted message. There
exist different ways of formalising the latter requirement as we
will see very soon below.

The task of private communication  over a noisy classical channel
has an old history. Wyner \cite{Wyner}, and Csisz\'{a}r and K\"{o}rner
\cite{CsiszarKorner} first studied this problem
for a point to point classical channel in the asymptotic setting
of many independent and identical (iid) uses of the channel. Calling
it the {\em wiretap} channel, they proved the following optimal bound:
\begin{equation}
\label{eq:cwiretap}
R_A = \max_{P} (I(X:C)_P - I(X:E)_P),
\end{equation}
where the channel $\mathfrak{C}$ is modelled as a stochastic map from 
set $A$ to set $C \times E$
and the mutual information is measured with respect to the 
probability distribution $p(x) p(c e | x)$ where $X$ is an auxilliary
random variable and 
the maximisation is done over all choices of the random variable
$X$ and encoding maps $x \mapsto p(a|x)$. The above bound is
obtained as follows. Let $n$ be the number of iid channel uses.
Alice chooses a random code book of size
$2^{n(R_A + r_a)}$ by independently sampling from the probability
distribution $p(x^n)$ on $X^n$. This code book is divided into 
$2^{n R_A}$ blocks, each of size $2^{n r_a}$. To transmit the $m$th 
message for an $m \in 2^{n R_A}$, Alice chooses a uniformly random codeword
from the $m$th block, say the $m'$th codeword $x^(m,m')$, applies
the stochastic encoding map to $x^n(m,m')$ to get
a sample from a probability distribution on $A^n$ and feeds it to $n$
copies of the channel $\mathfrak{C}$. The output of the channel is
a sample from the probability distribution $p(c^n e^n | x^n(m,m'))$.
Bob can decode the pair $(m,m')$ if the rate $R_A + r_a$ per channel
use is less than $I(X:C)_P$. On the other hand, Eve is `obfuscated' if
$r_a \geq I(X:E)_P$. This leads to the achievable rate 
$R_A \leq I(X:C)_P - I(X:E)_P$ for private classical communication. 
Here, different notions of secrecy lead to different notions of
formalisation of the statement `Eve is obfuscated'.

The multiple access channel (MAC) is arguably the simplest multiterminal
communication channel where there are several independent senders but 
only one
genuine receiver. Private communication over a MAC is an important
cryptographic task modelling, for example, the secure communication of 
messages from multiple independent agents in the field to a base station.
Here, the genuine receiver should be able to decode the entire transmitted
message tuple with small error and an eavesdropping receiver should
hardly get any information about the transmitted message tuple.
In the last decade several
authors have considered the problem of private classical communication
over various types of classical multiple access channels in the 
asymptotic iid setting culminating in the work of Chen, Koyluoglu
and Vinck \cite{CKV} who proved the following inner bound for a 
general classical discrete memoryless MAC in the asymptotic iid setting:
the union of rate regions of the form
\begin{equation}
\label{eq:cmac}
\begin{array}{rcl}
R_A & \leq & I(X:CY|Q)_P - I(X:E|Q)_P, \\
R_B & \leq & I(Y:CX|Q)_P - I(Y:E|Q)_P, \\
R_A + R_B & \leq & I(XY:C|Q)_P - I(XY:E|Q)_P,
\end{array}
\end{equation}
where the mutual information is measured with respect to the probability
distribution $p(q) p(x|q) p(y|q) p(c,e|x,y)$, $Q$ is an auxilliary
`time sharing' random variable, $X$, $Y$ are auxilliary random variables
that are independent given $Q$, $X \rightarrow A$, $Y \rightarrow B$
are independent stochastic encoding maps and the channel is a stochastic
map from $A \times B$ to $C \times E$. The union is taken over
all probability distributions of the form 
$p(q) p(x|q) p(y|q) p(a|x) p(b|y)$.

Both Equations\ref{eq:cwiretap} and \ref{eq:cmac} above use the 
asymptotically vanishing mutual information definition of secrecy
viz. they require that $I(M_A : E^n) / n$ or $I(M_A M_B: E^n) / n$  
approach zero as the
number of iid channel uses $n \rightarrow \infty$ where 
$(M_A, M_B)$ denote the input messages distributed uniformly in
$[2^{n R_A}] \times [2^{n R_B}]$. This definition is strictly
weaker than the small leakage in trace distance definition of secrecy
defined below that we will use in this paper. Nevertheless we will
be able to reproduce the above bounds even under the stronger secrecy
requirement.

The problem of private classical information over a point to point
quantum channel was first studied by Devetak \cite{Devetak} in the
asymptotic iid setting. The channel $\mathfrak{C}$ is 
modelled as a completely positive trace preserving (CPTP) map 
from density matrices on the input Hilbert space $A$ to 
density matrices on the output  Hilbert space $C \otimes E$. Here 
$A$ is the Hilbert spaces of a sender
Alice, $C$ is the Hilbert space of the genuine receiver Charlie
and $E$ is the Hilbert space of an eavesdropping receiver Eve. 
Devetak proved that
the natural regularised quantum analogue of Equation~\ref{eq:cwiretap}
is the optimal rate viz.
\begin{equation}
\label{eq:qwiretapiid}
R_A = \lim_{n \rightarrow \infty}
       n^{-1} \max_{\rho} (I(X:C^n)_\rho - I(X:E^n)_\rho),
\end{equation}
where the mutual information is taken over all classical quantum
states of the form
$
\rho^{X C^n E^n} = \sum_{x} p(x) \ket{X}^X \bra{x} \rho_x^{C^n E^n},
$
and the maximisation is done over all random variables $X$ and encoding
mappings $x \mapsto \sigma_x^{A^n}$. The state $\rho_x^{C^n E^n}$
is obtained by applying the channel $\mathfrak{C}^{\otimes n}$ to
$\sigma_x^{A^n}$. Subsequently Renes and Renner 
\cite{RenesRenner},
Radhakrishnan, Sen and Warsi \cite{rsw} and Wilde \cite{wildeWiretap} 
studied the quantum wiretap channel in the one shot setting culminating
in the optimal bound
\begin{equation}
\label{eq:qwiretaponeshot}
R_A = \max_{\rho} (I_H^\epsilon(X:C)_\rho - I_{\max}^\delta(X:E)_\rho),
\end{equation}
where the one shot mutual informations (defined formally later on)
are taken over all classical quantum
states of the form
$
\rho^{XCE} = \sum_{x} p(x) \ket{X}^X \bra{x} \rho_x^{CE},
$
and the maximisation is done over all random variables $X$ and encoding
mappings $x \mapsto \sigma_x^{A}$.  The state $\rho_x^{CE}$
is obtained by applying the channel $\mathfrak{C}$ to
$\sigma_x^{A}$. 
This one shot bound reduces to 
Devetak's bound in the asymptotic iid setting. 

The above works behoove us to study the one shot private classical
capacity of the quantum multiple access channel (QMAC). The channel 
$\mathfrak{C}$ is 
modelled as a CPTP map from input Hilbert space $A \otimes B$ to
output  Hilbert space $C \otimes E$. Here 
$A$, $B$ are to be thought of Hilbert spaces of two independent senders
Alice and Bob.
Alice gets a message $m_a \in [2^{R_A}]$ and Bob gets an independent 
message $m_b \in [2^{R_B}]$. Alice encodes $m_a$ into a density matrix
$\sigma_{m_a}^A$ in the Hilbert space $A$ and Bob independently encodes
$m_b$ into $\sigma_{m_b}^B$. Then $\sigma_{m_a}^A \otimes \sigma_{m_b}^B$
is fed into $\mathfrak{C}$ giving rise to a state 
$\rho_{m_a, m_b}^{CE}$ at the channel output. Let $0 < \epsilon, 
\delta < 1$. We require that, averaged over the uniform probability
distribution on $(m_a, m_b) \in [2^{R_A}] \times [2^{R_B}]$,
Charlie should be able to recover $(m_a, m_b)$ with probability
at least $\epsilon$ from $\rho_{m_a, m_b}^{C}$, and Eve's state
$\rho_{m_a, m_b}^{E}$ should be $\delta$-close to some fixed state
$\bar{\rho}^E$ in trace distance. We then say that 
$(R_A, R_B)$ is an achievable rate pair
for private classical communication over $\mathfrak{C}$ with
error $\epsilon$ and leakage $\delta$.
Note that Equation~\ref{eq:qwiretaponeshot} for the quantum wiretap channel
above holds for the stronger definition of leakage in 
trace distance, thus improving even on the classical asymptotic iid
wiretap results proved
earlier. The trace distance leakage definition is stronger because
$\delta_n$-leakage in trace distance implies asymptotically
vanishing mutual information leakage if $\delta_n \rightarrow 0$.
Continuing this tradition, in this paper we will aim for secrecy
in the trace distance leakage sense only.

It is thus natural to ponder about private classical communication over
a QMAC. A first attempt in this regard was made by Aghaee and Akhbari
\cite{AghaeeAkhbari} all the way in the one shot setting, but their 
proof has the following serious gap.
They use the single sender convex split lemma
of Anshu, Devabathini and Jain \cite{convexSplit} in order to guarantee 
individual secrecy for Alice and individual secrecy for Bob, but that 
does not guarantee joint
secrecy. A natural way to get joint secrecy would be to use 
the tripartite convex split lemma of Anshu, Jain and Warsi 
\cite{generalizedSlepianWolf} instead.
Indeed, Charlie can use the simultaneous QMAC decoder of Sen 
\cite{sen2020unions}
and Eve can be obfuscated via the tripartite convex split lemma in
order to get the following achievable rate region of private classical
communication over a QMAC: the union of rate regions of the form
\begin{equation}
\label{eq:qmacnonsmooth}
\begin{array}{rcl}
R_A & \leq & I_H^\epsilon(X:CY|Q)_\rho - I_{\max}(X:E|Q)_\rho, \\
R_B & \leq & I_H^\epsilon(Y:CX|Q)_\rho - I_{\max}(Y:E|Q)_\rho, \\
R_A + R_B & \leq & I_H^\epsilon(XY:C|Q)_\rho - I_{\max}(XY:E|Q)_\rho,
\end{array}
\end{equation}
where the mutual information is measured with respect to classical
quantum state of the form
\[
\rho^{QXYCE} =
\sum_{q,x,y} p(q) p(x|q) p(y|q)
\ket{q,x,y}^{QXY} \bra{q,x,y} \otimes \rho_{xy}^{CE},
\]
$Q$ is an auxilliary `time sharing' random variable, 
$X$, $Y$ are auxilliary random variables
that are independent given $Q$, $x \mapsto \sigma_x^A$, 
$y \mapsto \sigma_y^B$ are independent  encoding maps, and
the state $\rho_{xy}^{CE}$
is obtained by applying the channel $\mathfrak{C}$ to
$\sigma_x^{A} \otimes \sigma_y^B$. 
The union is taken over
all probability distributions of the form 
$p(q) p(x|q) p(y|q) p(a|x) p(b|y)$ and encoding maps
$x \mapsto \sigma_x^A$, $y \mapsto \sigma_y^B$.

Though we will not formally prove the achievability of 
Equation~\ref{eq:qmacnonsmooth} in this paper, the reader can easily
do so using the techiques outlined here combined with the tripartite
convex split lemma. However Equation~\ref{eq:qmacnonsmooth} has a 
big drawback viz. the terms for Eve are stated in 
terms of the non-smooth max mutual information even though the
terms for Charlie are stated in terms of the smooth hypothesis testing
mutual information. Because of this drawback, we cannot conclude that
in the asymptotic iid limit the one shot bounds lead to the natural
quantum version of Equation~\ref{eq:cmac}. The drawback
arises because the tripartite convex split lemma 
\cite{generalizedSlepianWolf} has only 
been proved
for non-smooth max mutual information. Proving it for smooth max
mutual information is related to the {\em simultaneous smoothing problem}
\cite{Drescher}, a major open problem in quantum information theory.

In this work, we obtain an alternate one shot
inner bound for private classical
communication over a QMAC that is stated in terms of smooth mutual
information quantities only. Our inner bound is contained inside the 
smooth version of the region of Equation~\ref{eq:qmacnonsmooth}.
Nevertheless we are able to show that in the asymptotic iid setting,
our one shot bound leads to the natural
quantum version of Equation~\ref{eq:cmac}. Our inner bound holds
for joint secrecy of Alice and Bob under the leakage in trace distance
definition and is the first non-trivial inner bound for 
private classical communication over a QMAC.

We prove our inner bound by using three powerful information
theoretic techniques. The first technique is the use of rate splitting,
originally developed by Grant et al. \cite{Rate_Splitting_Urbanke} in 
the classical
asymptotic iid setting, but recently extended to the one shot quantum
setting by the present authors \cite{biidi8}. Rate splitting allows
us to split one sender, say Alice, into two independent senders
Alice1 and Alice2. The two sender QMAC then becomes a three sender QMAC,
the advantage of which will become clear very soon. The second technique 
is simultaneous decoding
for sending classical information over a QMAC recently developed
by Sen \cite{sen2020unions}. Simultaneous decoding is used by 
Charlie to decode
Alice's message block and codeword within the block, which has been split 
into Alice1's and Alice2's parts,
and Bob's message block and codeword within the block, at any 
rate triple contained in the standard
polyhedral achievable region of a three sender MAC. The three senders also
have to ensure Eve's obfuscation which they do by randomising within a 
block as in
the proof of the original classical wiretap channel result of
Equation~\ref{eq:cwiretap}. The third and final technique that 
guarantees that this obfuscation strategy works is a novel result proved
in this paper called the {\em smoothed distributed covering lemma}.
This lemma is the main technical advancement of this work and should
be useful elsewhere. It is 
proved by repeated applications of the single sender convex split
lemma \cite{convexSplit}, which happens to hold for the smooth max mutual 
information. The lemma ensures the joint secrecy of Alice1, Alice2 and
Bob with a rate region described by smooth max mutual information 
quantitites. Though this region is inferior to what one would get from
a smoothed tripartite convex split lemma, it is nevertheless good
enough to lead to the desired region in the asymptotic iid setting.
The advantage of splitting Alice into Alice1 and Alice2 now becomes
clear because the split together with the distributed smoothed covering
lemma gives more obfuscation rate tuples. This leads to a larger inner
bound region for private classical communication than what one would
obtain otherwise without rate splitting. In particular the region obtained
without rate splitting seems to be insufficient to
obtain the desired rate region in the asymptotic iid limit in the
absence of a simultaneous smoothing result.

\section{Preliminaries}
All Hilbert spaces in this paper are finite dimensional. By 
$\mathcal{H}(A)$ we mean the Hilbert space associated with the system 
$A$. We will often use $\mathcal{H}(A)$ and $A$ interchangeably, in the 
sense that, when we say a state $\rho$ is defined on $A$, we mean the 
positive semidefinite matrix $\rho$ belongs to the Hilbert space 
$\mathcal{H}(A)$.

By the term `cq state' we mean some classical-quantum state $\rho^{XB}$ 
which is of the form

\begin{align*}
    \rho^{XB}\coloneqq \sum\limits_{x\in 
\mathcal{X}}\ket{x}\bra{x}^X\otimes \rho^B_x
\end{align*}

\begin{definition}
Let $\rho$ and $\sigma$ be two states in the same Hilbert space. Then, 
given $0\leq \epsilon <1$ we define the smooth hypothesis testing 
relative entropy of $\rho$ with respect to $\sigma$ aa
\begin{align*}
    D_H^{\epsilon}(\rho||\sigma)\coloneqq 
\max\limits_{\Pi:Tr[\Pi\rho]\geq 1-\epsilon}-\log \Tr[\Pi\sigma]
\end{align*}
\end{definition}
\begin{definition}
Given a state $\rho^{AB}$, the smooth hypothesis testing mutual 
information between $A$ and $B$ is defined as
\begin{align*}
    I_H^{\epsilon}(A:B)_{\rho}\coloneqq 
D_H^{\epsilon}(\rho^{AB}||\rho^A\otimes \rho^B)
\end{align*}
\end{definition}

We will require the notion of the \emph{purified distance}, which, for 
any two states $\rho$ and $\sigma$ in the space Hilbert space, is 
defined as
\begin{align*}
    P(\rho,\sigma)\coloneqq \sqrt{1-F^2(\rho,\sigma)}
\end{align*}
where $F(\rho,\sigma)$ is the fidelity between $\rho$ and $\sigma$. On 
occasion we will find it easier to use other metrics, such as the 
$1$-norm. To that end, the Fuchs-Van de Graaf inequalities essentially 
prove that all these metrics are equivalent:
\begin{fact}
For any two states $\rho$ and $\sigma$ in the same Hilbert space, the 
following holds
\begin{align*}
    1-\frac{1}{2}\norm{\rho-\sigma}_1\leq F(\rho,\sigma)\leq 
\sqrt{1-\frac{1}{4}\norm{\rho-\sigma}^2_1}
\end{align*}
\end{fact}

\begin{definition}\label{def:maxDivergence}
Given two states $\rho$ and $\sigma$ in the same Hilbert space, we 
define the max relative entropy of $\rho$ with respect to $\sigma$ as
\begin{align*}
    D_{\max}(\rho||\sigma) \coloneqq \inf \brak{\lambda~|~\rho\leq 
2^{\lambda}\sigma}
\end{align*}
\end{definition}

\begin{definition}
\label{def:smoothMaxDivergence}
Given the setting of \cref{def:maxDivergence}, the $\epsilon$ smooth max 
relative entropy is defined as
\begin{align*}
    D_{\max}^{\epsilon}(\rho||\sigma)\coloneqq \inf\limits_{\rho'\in 
B^{\epsilon}(\rho)} D_{\max}(\rho||\sigma)
\end{align*}
where $B^{\epsilon}(\rho)$ is the $\epsilon$ ball around $\rho$ with 
respect to the purified distance.
\end{definition}
\begin{definition}
Given a state $\rho^{AB}$, the smooth max mutual information between $A$ 
and $B$ is defined as
\begin{align*}
    I_{\max}^{\epsilon}(A:B)\coloneqq 
D_{\max}^{\epsilon}(\rho^{AB}||\rho^A\otimes \rho^B)
\end{align*}
\end{definition}

\section{Our results} 
We study the single shot private 
capacity of the classical quantum multiple access channel. The problem 
is as follows: we are given a quantum multiple access channel along with 
two independent classical distributions $P_X$ and $P_Y$ on the inputs 
for the two senders, Alice and Bob. Suppose that the input distributions 
are supported on the classical alphabets $\mathcal{X}$ and 
$\mathcal{Y}$. The output states corresponding to each input tuple 
$(x,y)$ is a shared quantum state between the receiver Charlie and the 
eavesdropper Eve. This situation is usually modelled by the following so 
called control state: 
\begin{align}\label{state:controlState}
    \rho^{XYCE}\coloneqq \sum\limits_{\substack{x\in \mathcal{X} \\ y 
\in \mathcal{Y}}}P_X(x)\cdot P_Y(y)\ket{x}\bra{x}^{X}\otimes 
\ket{y}\bra{y}^{Y}\otimes \rho_{x,y}^{CE} 
\end{align} 
The goal is for 
Alice and Bob to send messages $m$ and $n$ from the sets $[M]$ and $[N]$ 
via this channel to Bob in such a way that Eve does not gain any 
information about the message tuple that was sent, yet Charlie is able 
to decode both Alice an Bob's messages with high probability. To be 
precise, we require that, given $\epsilon, \delta>0$: 
\begin{enumerate}
    \item For all messages $m$ and $n$, 
    \begin{align*}
        \Pr[(\hat{m},\hat{n})\neq (m,n)]\leq \epsilon
    \end{align*}
    where $(\hat{m},\hat{n})$ is Charlie's estimate of the messages sent 
by Alice and Bob. This is called the \textbf{correctness} condition.
    \item There exists a state $\sigma^E$ such that, for all tuples $(m,n)$
    \begin{align*}
        \norm{\rho^E_{f(m),g(n)}-\sigma^E}_1\leq \delta
    \end{align*}
    where $\rho^E_{m,n}$ is the state induced on Eve's system when Alice 
and Bob send the messages $m$ and $n$ after encoding the messages into 
the input space of the channel via the maps $f: [M]\to \mathcal{X}$ and 
$g:[N]\to \mathcal{Y}$. This is called the \textbf{secrecy} condition. 
\end{enumerate}

\subsection{Previous Work}

A simpler variant of this problem, formally known as the 
classical-quantum wiretap channel, has been studied before in the 
one-shot setting by Radhakrishnan-Sen-Warsi \cite{rsw}. The heart of the 
argument used in that paper is a technical tool called the covering 
lemma. To gain some understanding of the RSW argument, consider the 
following strategy:

\begin{enumerate}
    \item Sender Alice chooses $2^{R}$ symbols 
$\brak{x_1,x_2,\ldots,x(2^R)}$ iid from her input distribution $P_X$.
    \item She then divides the list of $2^R$ symbols into blocks, each 
of size $2^K$.
    \item Alice then assigns a block number to each message $m\in [M]$.
    \item To send the message $m$, Alice first looks at the block of 
symbols corresponding to $m$, say $(x(i_1),x(i_2),\ldots,(i_{2^K}))$. 
She then \emph{randomly} picks an index $i_{\textsc{rand}}$ from this 
block and sends the corresponding symbol through the channel. 
\end{enumerate}

\textbf{Correctness :} It is known \cite{WangRenner} that as long as the 
rate $R-K$ is at most slightly less than the smooth hypothesis testing 
mutual information $I_H^{\epsilon}(X:C)$, the decoding error is at most 
$\epsilon$. Please note that the quantity $I_H^{\epsilon}(X:C)$ is 
computed with respect to the control state corresponding to only a 
single sender for this channel. \cite{AnshuJain}

\textbf{Secrecy :} To show that the secrecy condition holds, RSW proved 
a novel one-shot covering lemma. They showed that, as long as $K$ is 
slightly more than the smooth max mutual information 
$I_{\max}^{\delta}(X:E)$ (again computed with respect to the single 
sender control state), then, for every message $m\in [M]$, the following 
condition holds with high probability, over all choices of the codebook:
\begin{align*}
\norm{\frac{1}{K}\sum\limits_{j\in [K]}\rho^E_{i_j}-\rho^E}_1 \leq \delta
\end{align*}
where the indices $\brak{i_j}$ belong to the block corresponding to 
message $m$, and $\rho^E$ is the marginal of the control state on $E$.

To see that this implies that privacy holds in the protocol, notice that 
the expression on the right inside the norm is precisely the state 
induced by Alice's encoding function on the system $E$.

\subsubsection{The Single Shot Covering Lemma}

The covering lemma proved by RSW goes via an operator Chernoff bound. 
While this style of argument gives a strong concentration bound for the 
secrecy condition, one caveat is that the rate $K$ becomes dependant on 
the dimension of the eavesdropper system $E$. To be precise, for the 
secrecy condition to hold, RSW require the following condition:

\begin{align*}
    K \geq I_{\max}^{O(\delta)}(X:E)-\log \delta +\log\log\abs{E} +O(1)
\end{align*}

Strictly speaking, such a strong condition is not necessary to prove the 
covering lemma. One can show that the secrecy condition holds \emph{in 
expectation} over the choice of symbols inside the block. To make things 
precise, consider the following fact:

\begin{fact}
\label{fact:averageCovering}
Given the control state $\sum\limits_{x\in 
\mathcal{X}}P_X(x)\ket{x}\bra{x}^X\otimes \rho^{E}$ and $\delta>0$, let 
$\brak{x_1,x_2,\ldots, x_K}$ be iid samples from the distribution $P_X$. 
Then, given the condition

\begin{align*}
\log K \geq I_{\max}^{O(\delta)}(X:E)_{\rho}-\log\delta
\end{align*}
the following holds
\begin{align*}
    \E\limits_{x_1,x_2,\ldots,x_K}\norm{\frac{1}{K}\sum\limits_{i\in 
[K]}\rho^{E}_{x_i}-\rho^E}_1\leq \delta 
\end{align*} 
\end{fact}

This average version of the covering lemma is a direct consequence of 
the \emph{convex split lemma} proved by Anshu, Devabathini and Jain 
\cite{convexSplit}, adapted to cq states. A proof of 
\cref{fact:averageCovering} for the \emph{non-smooth} max information 
can be found in \cite{oneshotPrivacy}. The smoothing argument is 
standard and can be easily adapted from the smooth version if the convex 
split lemma proved by Wilde \cite{wildeWiretap}.

\subsection{Our Contribution}\label{sec:ourContribution}

As mentioned earlier we consider the problem of sending information 
privately over a classical-quantum multiple access channel in the single 
shot setting. The achievable rate region we would like to recover is as 
follows:

\begin{align*}
    &\log M \lesssim I_{H}^{\epsilon}(X:YC)-I_{\max}^{\delta}(X:E) \\
    &\log N \lesssim I_{H}^{\epsilon}(Y:XC)-I_{\max}^{\delta}(Y:E) \\
    & \log M+\log N \lesssim I_{H}^{\epsilon}(XY:C)-I_{\max}^{\delta}(XY:E)
\end{align*}

where we have omitted the $\log \epsilon$ and $\log \delta$ terms for 
clarity. To show that the above region is achievable, we will need the 
following technical tools:

\begin{enumerate}
    \item A \emph{distributed} covering lemma for multiple senders.
    \item A decoder which can decode any message pair, which corresponds 
to a rate in the desired region. 
\end{enumerate}

\subsubsection{A Smoothed Distributed Covering Lemma} 
We will address 
the second requirement later. For the distributed covering lemma, we 
wish to find the rate pairs $(K_1,K_2)$ such that, for $K_1$ and $K_2$ 
iid samples $\brak{x_1,x_2,\ldots, x_{K_1}}$ and $\brak{y_1,y_2,\ldots, 
y_{K_2}}$ from the distributions $P_X$ and $P_Y$ respectively, the 
following holds
\begin{align}
\label{eq:introSecurityCondition}
\E\limits_{\substack{x_1,x_2,\ldots,x_{K_1}\\ 
	             y_1,y_2,\ldots, y_{K_2}}}
\norm{\frac{1}{K_1\cdot K_2}
\sum\limits_{\substack{i\in [K_1] 
\\ j\in [K_2]}}\rho^{E}_{x_i,y_j}-\rho^E}_1\leq \delta
\end{align}

Notice that a na\"ive extension of the single user covering lemma will 
not work. This is because, the total number of random bits required for 
the secrecy condition is at least $I_{\max}^{\delta}(XY: E)$ bits while 
the na\"ive lemma would require only $\log K_1 + \log K_2 \geq 
I_{\max}^{\delta}(X: E)+I_{\max}^{\delta}(Y: E)$ random bits.

One way to prove the distributed covering lemma would be to appeal to a 
multipartite convex split lemma, and then exploit the connection between 
the convex split lemma for cq states and a covering lemma 
\cite{oneshotPrivacy}. Indeed such a \emph{non-smooth} multipartite 
version of the convex split lemma does exist and is not hard to prove 
\cite{generalizedSlepianWolf}. However, this proof strategy will give us 
a region of the following kind:

\begin{align*}
    &\log K_1 > I_{\max}(X:E)-\log\delta \\
    &\log K_2 > I_{\max}(Y:E)-\log\delta \\
    & \log K_1 + \log K_2 > I_{\max}(XY:E)-\log\delta
\end{align*}

One can see that this region is described in terms of the non-smooth max 
information. Indeed, obtaining the above region in terms of the smooth 
max information is a major open problem in quantum information theory, 
and is known as the \emph{simultaneous smoothing conjecture} 
\cite{Drescher}. In the absence of a smoothed region, we cannot hope to 
recover the desired rates in terms of the quantum mutual information in 
the asymptotic iid limit.

In this paper, we overcome this problem by taking a different approach. 
Instead of straightaway trying to show the secrecy property of the 
entire inverted pentagonal region (with two sides at infinity), we first 
prove a \emph{sequential} covering lemma for a corner point of the 
region. We show that, if Alice randomises over a block of size $\log K_1 
> I_{\max}^{\delta}(X:E)$ and Bob randomises over a block of size $\log 
K_2 > I_{\max}^{\delta}(Y: XE)$, then indeed 
\cref{eq:introSecurityCondition} holds, albeit with a worse dependence 
in $\delta$. A similar statement holds for the other corner point as 
well. We call this a \emph{successive cancellation} style covering 
lemma, since the strategy is similar in spirit to the successive 
cancellation style decoding for the multiple access channel.

To be precise, we prove the following lemma:
\begin{lemma}
\label{lem:succCancCoveringLemma}
Given the control state in \cref{state:controlState}, $\delta>0$ and 
$0<\epsilon'<\delta$ let $\brak{x_1,x_2,\ldots, x_{K_1}}$ and 
$\brak{y_1,y_2,\ldots, y_{K_2}}$ be iid samples from the distributions 
$P_X$ and $P_Y$. Then, if

\begin{align*}
    &\log K_1 \geq I_{\max}^{\delta-\epsilon'}(X:E)_{\rho}+\log 
\frac{3}{\epsilon'^3}-\frac{1}{4}\log\delta \\
    &\log K_2 \geq I_{\max}^{\delta-\epsilon'}(Y:EX)_{\rho}+\log 
\frac{3}{\epsilon'^3}-\frac{1}{4}\log\delta+O(1) 
\end{align*} 
the following holds 
\begin{align*}
 \E\limits_{\substack{x_1,x_2,\ldots,x_{K_1}\sim P_{X} \\ 
                      y_1,y_2,\ldots,y_{K_2}\sim P_{Y}}}
\norm{\frac{1}{K_1\cdot K_2}
\sum\limits_{i}^{K_2}\sum\limits_{j}^{K_1}\rho^E_{x_i, 
y_j}-\rho^E
}_1 \leq 20\delta^{1/8}
\end{align*}
\end{lemma}

\begin{remark} \label{remark;threeSender}

The proof of \cref{lem:succCancCoveringLemma} can be extended to the 
case when there are more than two senders. The argument is a 
straightforward induction on the triangle inequality in the last step of 
the proof. The dependence of the expected error on $\delta$ worsens 
however, with the constant increasing from $20$ to $40$ in the case when 
there are three senders.

\end{remark}

To recover the non-corner points in the idealised secrecy region, we use 
the idea of \emph{rate splitting}. Rate splitting was first suggested by 
Grant, Rimoldi, Urbanke and Whiting \cite{Rate_Splitting_Urbanke} as an 
alternative to time sharing to achieve the non-corner points on the 
dominant face of the achievable pentagon, in the context of sending 
classical information over a classical multiple access channel in the 
asymptotic iid setting. Recently, Chakraborty, Nema and Sen 
\cite{biidi8} adapted this technique to the one-shot 
fully quantum regime to derive entanglement transmission codes across a 
quantum multiple access channel.

The idea of rate splitting is roughly as follows : Given the input 
distribution $P_X$ corresponding to the sender Alice, we \emph{split} 
the distribution into two independent distributions $P_U^{\theta}$ and 
$P_V^{\theta}$, with respect to a parameter $\theta\in [0,1]$. These two 
new distributions correspond to two new senders Alice\sub{1} and 
Alice\sub{2}. $U^{\theta}$ and $V^{\theta}$ are independent random 
variables, each supported on the alphabet $\mathcal{X}$. This splitting 
is done by using a \emph{splitting function} $f:\mathcal{X}\times 
\mathcal{X}\to \mathcal{X}$, which has the following properties:

\begin{enumerate}
    \item $f(U^{\theta},V^{\theta})\sim P_X$ for all $\theta\in [0,1]$.
    \item For $\theta=0$ , $P_{f(U^{\theta},V^{\theta})|U^{\theta}}=P_X$ 
and for $\theta=1$,
    $P_{f(U^{\theta},V^{\theta})|U^{\theta}}$ puts all 
		its mass on one element.
    \item For a fixed $u$, $P_{f(U^{\theta},V^{\theta})|U^{\theta}}$ is 
a continuous function of $\theta\in [0,1]$. 
\end{enumerate}

Grant et.al. proved that such a family of triples 
$\brak{(P_U^{\theta},P_V^{\theta},f)}$ exists which obeys these 
properties. They did this via the following explicit construction:

For a fixed $\theta\in [0,1]$ and assuming that the elements of 
$\mathcal{X}$ have an ordering,
\begin{enumerate}
  \item $\Pr[U^{\theta}\leq u]\coloneqq \theta\cdot \Pr[X\leq u]+1-\theta$
    \item $\Pr[V^{\theta}\leq v]\coloneqq \frac{\Pr[X\leq 
v]}{\Pr[U^{\theta}\leq v]}$
    \item $f(u,v)\coloneqq \max(u,v)$
\end{enumerate}
We will refer to this construction as the max contruction.

Using this split, we can rewrite the control state in 
\cref{state:controlState} after splitting as follows:

\begin{align}\label{state:rateSplitControlState}
   \rho^{UVYCE}_{\theta}\coloneqq 
\sum\limits_{\substack{u,v\in \mathcal{X} \\ 
y \in \mathcal{Y}}}P_U^{\theta}(u)\cdot 
P_V^{\theta}(v)\cdot P_Y(y)\ket{u}\bra{u}^{U}\otimes 
\ket{v}\bra{v}^{V}\otimes \ket{y}\bra{y}^{Y}\otimes 
\rho(\theta)_{u,v,y}^{CE}
\end{align}
where for each $(u,v)$
\begin{align*}
    \rho(\theta)_{u,v,y}^{CE}\coloneqq \rho_{f(u,v),y}^{CE}
\end{align*}

Armed with this split state, we invoke the three sender version of 
\cref{lem:succCancCoveringLemma} to prove the following theorem:
\begin{theorem}
\label{thm:mainSuccCanccCovering}
Given the control state in \cref{state:rateSplitControlState}, 
$\delta>0$ and $0<\epsilon'<\delta$ let $\brak{u_1,u_2,\ldots, 
u_{K_1}}$, $\brak{y_1,y_2,\ldots, y_{K_2}}$ and $\brak{v_1,v_2,\ldots, 
v_{K_3}}$ be iid samples from the distributions $P_U^{\theta}$, $P_Y$ 
and $P_V^{\theta}$ respectively, for a fixed $\theta\in [0,1]$. Then, if

\begin{align*}
    &\log K_1 \geq 
I_{\max}^{\delta-\epsilon'}(U^{\theta}:E)_{\rho_{\theta}}+\log 
\frac{3}{\epsilon'^3}-\frac{1}{4}\log\delta \\
    &\log K_2 \geq 
I_{\max}^{\delta-\epsilon'}(Y:EU^{\theta})_{\rho_{\theta}}+\log 
\frac{3}{\epsilon'^3}-\frac{1}{4}\log\delta+O(1) \\
    &\log K_3 \geq 
I_{\max}^{\delta-\epsilon'}(V^{\theta}:EYU^{\theta})_{\rho_{\theta}}+\log 
\frac{3}{\epsilon'^3}-\frac{1}{4}\log\delta+O(1)
\end{align*}
the following holds
\begin{align*}
    \E\limits_{\substack{u_1,u_2,\ldots, u_{K_1}\sim P_{U}^{\theta} \\ 
y_1,y_2,\ldots,y_{K_2}\sim P_{Y}\\ v_1,v_2,\ldots, v_{K_3}\sim 
P_V^{\theta}}}\norm{\frac{1}{K_1\cdot K_2\cdot 
K_3}\sum\limits_{i,j,k}^{K_1,K_2,K_3}\rho(\theta)^E_{u_i, 
y_j,v_k}-\rho^E}_1 \leq 40\delta^{1/8}
\end{align*}
\end{theorem}

\begin{remark}
~\begin{enumerate}
    \item Note that by construction of the triple 
$(P_U^{\theta},P_V^{\theta},f)$, 
\begin{align*}\rho^E=\rho^E_{\theta}\end{align*}
    \item Alice has to randomise over of \emph{total} block of size of 
$K_1\cdot K_3$. This implies that, thinking of Alice as the combination 
of the two senders Alice\sub{1} and Alice\sub{2}, the size of the block 
over which Alice has to randomize has to be at least
    \begin{align*}
I_{\max}^{\delta-\epsilon'}(U^{\theta}:E)_{\rho_{\theta}}+
I_{\max}^{\delta-\epsilon'}(V^{\theta}:EYU^{\theta})_{\rho_{\theta}}+
2\log \frac{3}{\epsilon'^3}-\frac{1}{2}\log\delta+O(1)
    \end{align*}

    \item For $\theta=0$ and $\theta=1$, the expressions 
$,I_{\max}^{\delta-\epsilon'}(U^{\theta}:E)_{\rho_{\theta}}$ and 
$I_{\max}^{\delta-\epsilon'}(V^{\theta}:EYU^{\theta})_{\rho_{\theta}}$ 
take the value zero respectively. This can be easily seen from the 
properties of the max construction.
    \item When $\theta\in \brak{0,1}$, the secrecy region collapses to 
the two sender case. For $\theta=0$, the user Alice\sub{1} becomes 
trivial, and similarly for Alice\sub{2} when $\theta=1$. These values of 
$\theta$ thus correspond to the corner points of the secrecy region.

\end{enumerate}
\end{remark}

As $\theta$ ranges from $0$ to $1$, the point 
$(I_{\max}^{\delta-\epsilon'}(U^{\theta}:E)+
I_{\max}^{\delta-\epsilon'}(V^{\theta}:EYU^{\theta}), 
I_{\max}^{\delta-\epsilon'}(Y:EU^{\theta}))$ traces out a curve between 
the corner points, which lies on or above the line joining the corner 
points. To show that this is true, we use the following properties of 
the smooth max mutual information:

\begin{lemma}
\label{lem:invarianceLemma}

Given the control state in \cref{state:controlState} and the post split 
state in \cref{state:rateSplitControlState} for some fixed $\theta\in 
[0,1]$, the following holds

\begin{align*}
    I_{\max}^{\epsilon}(U^{\theta}V^{\theta}Y:E)_{\rho_{\theta}}=
I_{\max}^{\epsilon}(XY:E)_{\rho}
\end{align*}
for any $\epsilon>0$.
\end{lemma}

\begin{lemma}\label{lem:chainRule} 

Given a state $\varphi^{RAB}$, not necessarily pure, and $\epsilon>0$, 
the following holds

\begin{align*}
    I_{\max}^{12\epsilon}(R:AB)_{\varphi} \leq 
I_{\max}^{\epsilon-\gamma}(R:A)_{\varphi}+
I_{\max}^{\epsilon-\gamma}(RA:B)_{\varphi}+
2\log\frac{1}{\epsilon}+\log\frac{3}{\gamma^2}
\end{align*}
\end{lemma}

These two lemmas together show that the boundary of the secrecy region 
between the corner points lies on or above the straight line 
$x+y=I_{\max}^{O(\epsilon)}(XY:E)_{\rho}$.

\subsubsection{Decoding}

We now turn our attention to the problem of Charlie decoding the 
messages sent by Alice and Bob. There are two kinds of decoders we can 
consider:

\begin{enumerate}
    \item \textbf{Successive Cancellation:} One way to decode the 
messages would be a successive cancellation strategy, in which Charlie 
first decodes Alice\sub{1}, then using Alice\sub{1}'s message as side 
information he decodes Bob, and finally using Alice\sub{1} and Bob's 
messages as side information he decodes Alice\sub{2}. This gives us an 
achievable region which is the union over $\theta\in [0,1]$ over all 
rectangles subtended by the point 
\[(
I_{H}^{\epsilon}(U^{\theta}:C)+I_{H}^{\epsilon}(V^{\theta}:CU^{\theta}Y), 
I_{H}^{\epsilon}(Y:CU^{\theta}))
\]
    where we have neglected the additive $\log\epsilon$ terms for brevity.
    
    \begin{remark}

    For reasons that will become clear shortly, instead of following the 
order of decoding given above, we actually would like to decode in the 
order Alice\sub{1}--Bob--Alice\sub{1}. This would give the rate point

\[
(I_{H}^{\epsilon}(V^{\theta}:C)+
I_{H}^{\epsilon}(U^{\theta}:CV^{\theta}Y), I_{H}^{\epsilon}(Y:CV^{\theta}))
\]

    \end{remark}

    \item \textbf{Simultaneous Decoding:} The other decoding strategy we 
consider is simultaneous decoding. Given a cq-mac and the control state 
in \cref{state:controlState}, a simultaneous decoder gives us the 
following achievable region:

\begin{align*}
   & R_1< I_{H}^{\epsilon}(X:YC)-\log\frac{1}{\epsilon} \\
   & R_2< I_{H}^{\epsilon}(Y:XC)-\log\frac{1}{\epsilon} \\
   & R_1+R_2< I_{H}^{\epsilon}(XY:C)-\log\frac{1}{\epsilon} \\
\end{align*}
where $R_1$ and $R_2$ correspond to Alice and Bob's rates.
\end{enumerate} 

 In the absence of chain rules for the smooth hypothesis testing mutual 
information, the rate region given by the successive cancellation 
decoder is a deformed version of the pentagonal region we expect.

%


This issue can be mitigated somewhat if we use the simultaneous decoder. 
Thus, we will use simultaneous decoding as our decoding strategy of 
choice. We elaborate on this in the next section.

\subsubsection{Simultaneous Decoding}

We will use the construction given by Sen in \cite{sen2020unions}. Until 
recently, the existence of such a simultaneous decoder for the cq-mac 
which recovers the rate region given by Winter in \cite{winterMAC} in 
the asymptotic iid setting was a major open problem. To be precise, Sen 
proved the following fact:

\begin{fact}\label{fact:pranablemma}

Given a cq mac and its associated control state 
\cref{state:controlState}, there exists an encoding and decoding scheme 
such that, all rate pairs $(R_1,R_2)$, where $R_1$ corresponds to Alice 
and $R_2$ corresponds to Bob, are achievable for transmission of 
classical information of the channel with error at most 
$49\sqrt{\epsilon}$ :

\begin{equation*}
    \boxed{
    \begin{aligned}
    &R_1\leq I_H^{\epsilon}(X:YC)+\log \epsilon-1 \\
    &R_2\leq I_H^{\epsilon}(Y:XC)+\log \epsilon-1 \\
    &R_1+R_2 \leq I_H^{\epsilon}(XY:C)+\log \epsilon-1
\end{aligned}
}
\end{equation*}
\end{fact}

The above lemma is easily generalised to the case when there are 
multiple senders. In our case, we use a three sender simultaneous 
decoder, which, for every fixed $\theta\in [0,1]$ gives us the following 
achievable region for Alice\sub{1}, Bob and Alice\sub{2}:

\begin{equation}
\label{eq:rateRegion}
    \boxed{
    \begin{aligned}
    \nonumber&R_{10}< 
I_H^{\epsilon}(U^{\theta}:CV^{\theta}Y)+\log\epsilon-1 \\ \nonumber
    &R_2< I_H^{\epsilon}(Y:CU^{\theta}V^{\theta})+\log\epsilon-1 \\
    \nonumber&R_{11}< 
	   I_H^{\epsilon}(V^{\theta}:CU^{\theta}Y)+\log\epsilon-1 \\
   \nonumber &R_{10}+R_{11} < 
I_H^{\epsilon}(U^{\theta}V^{\theta}:CY)+\log\epsilon-1 \\
   \nonumber &R_{10}+R_2 < 
I_H^{\epsilon}(U^{\theta}Y:CV^{\theta})+\log\epsilon-1 \\
    \nonumber&R_2+R_{11} < 
I_H^{\epsilon}(YV^{\theta}:CU^{\theta})+\log\epsilon-1 \\
    &R_{10}+R_2+R_{11} < 
	   I_H^{\epsilon}(U^{\theta}YV^{\theta}:C)+\log\epsilon-1
\end{aligned}
}
\end{equation}

Here $R_{10}, R_2$ and $R_{11}$ corresponds to Alice\sub{1}, Bob and 
Alice\sub{2} respectively. The bound on the last term is equal to 
$I_H^{\epsilon}(XY:C)+\log\epsilon-1$. The proof of this fact is the 
same as \cref{lem:invarianceLemma}.

For every $\theta\in [0,1]$, we will project the above rate region to 
the $2$ dimensional space which contains the achievable rate points for 
Alice and Bob. To obtain the full achievable region, we take a union 
bound over all $\theta$. to be precise, we show the following lemma:

\begin{lemma}
\label{lem:3senderMac}

Given a $2$ sender cq mac and the associated control state in 
\cref{state:controlState}, and its corresponding split state 
\cref{state:rateSplitControlState} for some fixed $\theta\in [0,1]$, the 
following rate region is achievable for sending classical information 
over the channel with error $\epsilon^{1/8}$ is as follows:

\begin{equation}
\boxed{
\begin{aligned}
    &R_1 \leq I_H^{\epsilon}(U^{\theta}V^{\theta}:CY)+\log\epsilon-1 \\
    & R_1 \leq 
I_H^{\epsilon}(V^{\theta}:CU^{\theta}Y)+
I_H^{\epsilon}(U^{\theta}:CV^{\theta}Y)+2\log\epsilon-2 
\\
    & R_2 \leq I_H^{\epsilon}(Y:CU^{\theta}V^{\theta})+\log\epsilon-1 \\
    & R_2\leq I_H^{\epsilon}(YV^{\theta}:CU^{\theta})+\log\epsilon-1 \\
    & R_2 \leq I_H^{\epsilon}(U^{\theta}Y:CV^{\theta})+\log\epsilon-1 \\
    &R_1+R_2 \leq 
I_H^{\epsilon}(V^{\theta}:CU^{\theta}Y)+
	I_H^{\epsilon}(U^{\theta}Y:CV^{\theta})+2\log\epsilon-2 
\\
    & R_1+R_2 \leq I_H^{\epsilon}(YV^{\theta}:CU^{\theta}) + 
I_H^{\epsilon}(U^{\theta}:CV^{\theta}Y)+2\log\epsilon-2 \\
    &R_1+2R_2 \leq 
I_H^{\epsilon}(U^{\theta}Y:CV^{\theta})+
	I_H^{\epsilon}(YV^{\theta}:CU^{\theta})+2\log\epsilon-2 
\\
    &R_1+R_2 \leq I_H^{\epsilon}(U^{\theta}YV^{\theta}:C)+\log\epsilon-1
\end{aligned}
}
\end{equation}
where all the mutual information terms are computed with respect to 
\cref{state:rateSplitControlState}.
\end{lemma}

\subsubsection{The Private Capacity Region}

Let us call the achievable region given by \cref{lem:3senderMac} as 
$\mathcal{S}_{\theta}$, for some fixed $\theta\in [0,1]$. For the same 
$\theta$, consider the block sizes $(K_1+K_3,K_2)$ given by 
\cref{thm:mainSuccCanccCovering}. We define

\[
\mathcal{T}_{\theta}\coloneqq \brak{(\log K, \log K')~|~K\geq K_1\cdot 
K_3, K'\geq K_2}
\]

Then, we can have the following theorem, which gives an inner bound on 
the region for private transmission of classical information over the cq 
mac

\begin{theorem}\label{thm:mainPrivateRegion}

Given a classical quantum multiple access channel, the control state in 
\cref{state:controlState} and a split $(P_U^{\theta},P_V^{\theta},f)$ of 
the distribution $P_X$, for some $\theta\in [0,1]$ the rate pairs in the 
following region, are achievable for private transmission of messages 
across the channel
\begin{align*}
    \left (\bigcup\limits_{\theta\in [0,1]}\Big( \mathcal{S}_{\theta}- 
\mathcal{T}_{\theta} \Big)\right)^{+}
\end{align*}
with decoding error at most $49\sqrt{\epsilon}$ and privacy leakage at 
most $40\delta^{1/8}$, where $\epsilon, \delta>0$ and $0<\epsilon< 
\delta$. All the information quantities above are computed with respect 
to the split state $\rho_{\theta}^{UVYCE}$. Here, the operation 
$(A-B)^+$ , where $A$ and $B$ are sets of real numbers is defined as 
$\brak{\max(a-b,0)~|~a\in A, b\in B}$.
\end{theorem}
To precisely describe the set $\mathcal{S}_{\theta}-\mathcal{T}_{\theta}$, let
\begin{align*}
    & \delta' \coloneqq \delta-\epsilon' \\
    & c\coloneqq\log \frac{1}{\epsilon}+ 
	\log\frac{1}{\epsilon'^3}-\frac{1}{4}\log \delta  + O(1)
\end{align*}
and define the rates
\begin{align*}
   & R_A \coloneqq R_1-\log K \\
   & R_B \coloneqq R_2 -\log K'
\end{align*}
To ease the burden on notation, we drop the superscripts from the 
random variables $U^{\theta}$ and $V^{\theta}$. Then, for a fixed 
$\theta\in [0,1]$, the region 
$\mathcal{S}_{\theta}-\mathcal{T}_{\theta}$ looks like 
\begin{equation*}
    \boxed{
    \begin{aligned}
    &R_A \leq I_H^{\epsilon}(UV:YC)-
	    I_{\max}^{\delta'}(U:E)-I_{\max}^{\delta'}(V:UYE)+c \\
    & R_A \leq I_H^{\epsilon}(V:UYC)+I_H^{\epsilon}(U:VYC)-
	    I_{\max}^{\delta'}(U:E)-I_{\max}^{\delta'}(V:UYE)+2c \\
    & R_B \leq I_H^{\epsilon}(Y:UVC)-I_{\max}^{\delta'}(Y:UE)+c \\
    & R_B\leq I_H^{\epsilon}(YV:UC)-I_{\max}^{\delta'}(Y:UE)+c \\
    & R_B \leq I_H^{\epsilon}(UY:VC)-I_{\max}^{\delta'}(Y:UE)+c \\
    &R_A+R_B \leq I_H^{\epsilon}(V:UYC)+I_H^{\epsilon}(UY:VC)-
	    I_{\max}^{\delta'}(U:E)-I_{\max}^{\delta'}(V:UYE)-
	    I_{\max}^{\delta'}(Y:UE)+2c \\
    & R_A+R_B \leq I_H^{\epsilon}(YV:UC) + I_H^{\epsilon}(U:VYC)-
	    I_{\max}^{\delta'}(U:E)-I_{\max}^{\delta'}(V:UYE)-
	    I_{\max}^{\delta'}(Y:UE)+2c \\
    &R_A+2R_B \leq I_H^{\epsilon}(UY:CV)+I_H^{\epsilon}(YV:UC)-
	    I_{\max}^{\delta'}(U:E)-I_{\max}^{\delta'}(V:UYE)-
	    2I_{\max}^{\delta'}(Y:UE)+2c \\
    &R_A+R_B \leq I_H^{\epsilon}(UYV:C)-I_{\max}^{\delta'}(U:E)-
	    I_{\max}^{\delta'}(V:UYE)-I_{\max}^{\delta'}(Y:UE)+c
\end{aligned}
}
\end{equation*}

\begin{remark} 
\begin{enumerate}
    \item To get an idea as to what the above region looks like, first 
note that for $\theta=\brak{0,1}$, the region in \cref{lem:3senderMac} 
is equivalent to the following region:
    \begin{align*}
        &R_1\leq I_H^{\epsilon}(X:CY)+O(\log \epsilon) \\
        &R_2 \leq I_H^{\epsilon}(Y:CX) +O(\log \epsilon)\\
        &R_1+R_2 \leq I_H^{\epsilon}(XY:C)+O(\log \epsilon)
    \end{align*}

    \item This essentially looks like the achievable rate region for the 
$2$-sender mac. In fact, from \cref{lem:3senderMac} we can see that as 
$\theta$ ranges from $0$ to $1$, the corresponding rate regions 
$\mathcal{S}_{\theta}$ that we get are subsets of this pentagonal 
region.

    \item Note that, if the smooth hypothesis testing mutual information 
obeyed a chain rule with equality, then the region in 
\cref{lem:3senderMac} would be equivalent to the pentagonal region in 
Item 1 for \emph{all} values of $\theta\in [0,1]$.

    \item On the other hand, the secrecy region given by 
\cref{thm:mainSuccCanccCovering} and following it, looks like an 
inverted pentagon in the first quadrant with two sides at infinity, and 
the dominant face slightly warped due the chain rule for the smooth max 
information \cref{lem:chainRule}.

    \item The final secrecy region thus looks like a smaller pentagon, 
but with the dominant face warped inwards.

\end{enumerate}
\end{remark}

\subsection{Extension to the Asymptotic IID Regime}

In this we show that our one-shot techniques can be used to recover the 
expected private capacity region of the classical quantum multiple 
access channel in the limit of asymptotically many channel uses. To do 
this, we first note some facts about the asymptotic behaviour of the 
smoothed information quantities we have used so far:

\begin{fact}
\label{lem:asymIID}
Given a classical quantum state with $N$ classical inputs
\begin{align*}
    \rho^{X_1X_2\ldots X_NC}\coloneqq \sum\limits_{x_1x_2\ldots 
x_N}\prod\limits_{i}^{N}P_{X_i}(x_i)\ket{x_i}\bra{x_i}^{X_i}\otimes 
\rho_{x_1x_2\ldots x_N}^{C}
\end{align*}
let $J\subseteq [N]$. Then, for some $\epsilon>0$ and an integer $n\in 
\mathbb{N}$, the following holds true in the limit of $n\to \infty$ and 
$\epsilon\to 0$ for all $J$,

\begin{align*}
    &\lim\limits_{\epsilon\to 0}\lim\limits_{n\to 
\infty}\frac{1}{n}I_H^{\epsilon}(X_J^n:C^nX_{J^c}^n)_{\rho^{\otimes n}}= 
I(X_J:CX_{J^c})_{\rho} \\
    &\lim\limits_{\epsilon\to 0}\lim\limits_{n\to 
\infty}\frac{1}{n}I_{\max}^{\epsilon}(X_J^n:C^nX_{J^c}^n)_{\rho^{\otimes 
n}}= I(X_J:CX_{J^c})_{\rho}
\end{align*}
where $X_J\coloneqq \prod\limits_{j\in J}X_{j}$.
\end{fact}
\cref{lem:asymIID} allows us to prove the following theorem from 
\cref{thm:mainPrivateRegion}:

\begin{theorem}\label{thm:asymIID}

Given a classical quantum multiple access channel, the control state in 
\cref{state:controlState}, the following rate region is achievable for 
private transmission of messages across the channel, when asymptotically 
many channel uses are allowed:
\begin{align*}
    &R_{\textsc{Alice}}^{private} < I(X:YC)-I(X:E) \\
    &R_{\textsc{Bob}}^{private} < I(Y:XC)-I(Y:E) \\
    &R_{\textsc{Alice}}^{private}+ R_{\textsc{Bob}}^{private} < 
	I(XY:C)-I(XY:E)
\end{align*}
\end{theorem}

\subsection{A Generalisation}

A generalisation of the theorems presented in the previous sections can 
be shown to be true using a time sharing random variable. To be precise, 
instead of the input distributions $P_X$ and $P_Y$ on the classical 
alphabets $\mathcal{X}$ and $\mathcal{Y}$, we will consider the joint 
distribution $P_Q\otimes P_{X|Q}\cdot P_{Y|Q}$ over the alphabet 
$\mathcal{Q}\times \mathcal{X}\times \mathcal{Y}$. Consider the control 
state

\begin{align}
\label{state:QcontrolState}
        \rho^{QXYCE}\coloneqq \sum\limits_{q,x,y}P_Q(q)P_{X|Q}(x|q)\cdot 
P_{Y|Q}(y|q)\ket{q,x,y}\bra{q,x,y}^{QXY}\otimes \rho_{x,y}^{CE}
\end{align}
Define 
\begin{align*}
    \rho^{XYCE|Q}\coloneqq \big(\rho^Q\otimes 
\mathbb{1}\big)^{-1}\frac{\rho^{QXYCE}}{\text{rank}(\rho^Q)} 
\big(\rho^Q\otimes \mathbb{1}\big)^{-1}
\end{align*}

Using the above state, one can define the \emph{conditional} smooth 
hypothesis testing mutual information and the smooth max information. A 
version of \cref{fact:pranablemma} with respect to the above conditional 
control state was shown to be true in \cite{sen2020unions}. It is also 
not hard to see that the successive cancellation covering lemma 
\cref{lem:succCancCoveringLemma}, can also be proved using this control 
state, since the operator inequalities used in the proof of that lemma 
are preserved by the above definition.

Before we go on to state the general theorem with respect to the state 
\cref{state:QcontrolState}, we would like to remark that in order to get 
the most general version of the private capacity region, we consider a 
fully quantum or \emph{qq} multiple access channel $\mathfrak{C}$ which 
maps the systems $X'Y'\to CE$. To import this into the classical quantum 
setting, we introduce the classical alphabets $\mathcal{X}$ and 
$\mathcal{Y}$ and the maps $\mathfrak{F}:\mathcal{X}\to X'$ and 
$\mathfrak{G}:\mathcal{Y}\to Y'$ such that

\begin{align*}
    &\mathfrak{F}(x)\coloneqq \sigma^{X'}_x \\
    &\mathfrak{G}(y)\coloneqq \sigma^{Y'}_y
\end{align*}

where $\sigma^{X'}_x$ and $\sigma^{Y'}_y$ are states in the input 
Hilbert space of $\mathfrak{C}$.

Then define
\begin{align}\label{def:qqMapCq}
    \mathfrak{C}(\sigma^{X'}_x\otimes\sigma^{Y'}_y)\coloneqq 
	\rho^{CE}_{x,y}
\end{align}
We are now ready to state the theorem:
\begin{theorem}\label{thm:QprivateRegion}

Given the channel $\mathfrak{C}:X'Y'\to CE$, the maps $\mathfrak{F}, 
\mathfrak{G}$ and the definition \cref{def:qqMapCq}, consider the 
classical quantum control state given in \cref{state:QcontrolState}. 
Then, given the split $(P_{U|Q}^{\theta},P_{V|Q}^{\theta},f)$ with 
respect to the parameter $\theta\in [0,1]$, we have that the following 
region if achievable for private information transmission with error at 
most $\epsilon^{1/8}$ and leakage at most $40\delta^{1/8}$
\begin{equation*}
\boxed{
\begin{aligned}
    &R_A \leq I_H^{\epsilon}(UV:YC|Q)-I_{\max}^{\delta'}(U:E|Q)-
	 I_{\max}^{\delta'}(V:UYE|Q)+c \\
    & R_A \leq I_H^{\epsilon}(V:UYC|Q)+I_H^{\epsilon}(U:VYC|Q)-
	I_{\max}^{\delta'}(U:E|Q)-I_{\max}^{\delta'}(V:UYE|Q)+2c \\
    & R_B \leq I_H^{\epsilon}(Y:UVC|Q)-I_{\max}^{\delta'}(Y:UE|Q)+c \\
    & R_B\leq I_H^{\epsilon}(YV:UC|Q)-I_{\max}^{\delta'}(Y:UE|Q)+c \\
    & R_B \leq I_H^{\epsilon}(UY:VC|Q)-I_{\max}^{\delta'}(Y:UE|Q)+c \\
    &R_A+R_B \leq I_H^{\epsilon}(V:UYC|Q)+I_H^{\epsilon}(UY:VC|Q)-
	   I_{\max}^{\delta'}(U:E|Q)-I_{\max}^{\delta'}(V:UYE|Q)-
	   I_{\max}^{\delta'}(Y:UE|Q)+2c \\
    & R_A+R_B \leq I_H^{\epsilon}(YV:UC|Q) + I_H^{\epsilon}(U:VYC|Q)-
	I_{\max}^{\delta'}(U:E|Q)-I_{\max}^{\delta'}(V:UYE|Q)-
	I_{\max}^{\delta'}(Y:UE|Q)+2c \\
    &R_A+2R_B \leq I_H^{\epsilon}(UY:CV|Q)+I_H^{\epsilon}(YV:UC|Q)-
	  I_{\max}^{\delta'}(U:E|Q)-I_{\max}^{\delta'}(V:UY|QE)-
	  2I_{\max}^{\delta'}(Y:UE|Q)+2c \\
    &R_A+R_B \leq I_H^{\epsilon}(UYV:C|Q)-I_{\max}^{\delta'}(U:E|Q)-
	  I_{\max}^{\delta'}(V:UYE|Q)-I_{\max}^{\delta'}(Y:UE|Q)+c
\end{aligned}
}
\end{equation*}
and
\begin{equation*}
    \boxed{
    \begin{aligned}
    & \delta' \coloneqq \delta-\epsilon' \\
    & c\coloneqq\log \frac{1}{\epsilon}+ \log\frac{1}{\epsilon'^3}-
	      \frac{1}{4}\log \delta  + O(1)
    \end{aligned}
    }
\end{equation*}
where $\epsilon, \delta>0$, $0<\epsilon'<\delta$ and all the information 
quantities are computed with respect to the split of the control state 
in \cref{state:QcontrolState}.
\end{theorem}

\section{Proofs of Important Lemmas}
In this section we present the proof of all the lemmas and theorems 
stated in \cref{sec:ourContribution}.
\begin{proof}[Proof of \cref{lem:succCancCoveringLemma}]
Suppose we are given the cq state 
\begin{align*}
\rho^{XYE}\coloneqq \sum\limits_{\substack{x\in \mathcal{X} \\ y\in 
\mathcal{Y}}}p_X(x)p_Y(y)\ket{x}\bra{x}^{X}\otimes \ket{y}\bra{y}^{Y} 
\rho^{E}_{x,y}
\end{align*}

Consider the quantity
\begin{align*}
\lambda\coloneqq \imaxt(Y:EX)_{\rho} \coloneqq 
\inf\limits_{\norm{\rho'-\rho}_1\leq 
\epsilon}D_{\max}(\rho'^{XYE}||\rho^{Y}\otimes \rho'^{XE})
\end{align*}
Let $\tilde{\rho}^{XYE}$ be the optimizer in the definition of 
$\imaxt(Y:EX)$. Without loss of generality we can assume that 
$\tilde{\rho}^{XYE}$ is a cq state. This is because, suppose the 
optimizer was a state $\rho^*$ which is not cq. By definition, $\rho^*$ 
obeys the following properties
\begin{align*}
&\rho^* \leq 2^{\imaxt(Y:EX)}~\rho^{Y}\otimes \rho^{*XE} \\
    &\norm{\rho^*-\rho}_1\leq \epsilon
\end{align*}

Now we will measure the $X$ and $Y$ systems along the canonical bases 
$\brak{\ket{x}}$ and $\brak{\ket{y}}$, to get the cq state $\rho^{**}$. 
Since measurement is CPTP it preserves the operator inequality. This 
also implies that $\rho^{**}$ is in the $\epsilon$ ball around $\rho$. 
Finally, it is easy to see that, $\rho^{**XE}$ is the post measurement 
state on the systems $XE$. These observations imply that $\rho^{**}$ is 
a cq state which is also an optimizer, proving the claim.

Next, suppose that 
\begin{align*}
    \tilde{\rho}\coloneqq 
\sum\limits_{x,y}\tilde{P}_{XY}(x,y)~x^X\otimes y^Y\otimes 
\tilde{\rho}^{E}_{x,y}
\end{align*}
where we have used the shorthand $x^X\coloneqq \ket{x}\bra{x}^X$ and 
similarly for $y$. It is easy to see that the two following properties 
hold

\begin{align}
    &\norm{\tilde{P}_{XY}-P_X\cdot P_Y}_1 \leq \epsilon 
\label{eq:diffTildeOriginal}
\end{align}

\textbf{Changing the Distributions}
We can infer from \cref{eq:diffTildeOriginal} that 
\begin{align*}
    \norm{\Tilde{P}_x-P_X}_1 \leq \epsilon
\end{align*}
which implies that 
\begin{align}\label{eq:diffTildeOriginal2}
    \norm{\tilde{P}_{XY}-\Tilde{P}_X\cdot P_Y}_1 \leq 2\epsilon
\end{align}
\cref{eq:diffTildeOriginal2} can be written as 
\begin{align*}
    \E\limits_{\Tilde{P}_X\cdot 
P_Y}~\left[~\left\lvert\frac{\Tilde{P}_{XY}(X,Y)}{\Tilde{P}_X(X)\cdot 
P_Y(Y)}-1\right\rvert~\right]\leq 2\epsilon
\end{align*}
Then, by Markov's inequality this implies that 
\begin{align}\label{eq:changinDistributions}
    \Pr_{\Tilde{P}_X\cdot 
P_Y}\left[~\left\lvert\frac{\Tilde{P}_{XY}(X,Y)}{\Tilde{P}_X(X)\cdot 
P_Y(Y)}-1\right\rvert\geq \sqrt{\epsilon}~\right] \leq 2\sqrt{\epsilon}
\end{align}

Now, by the definition of $\Tilde{\rho}$ and using the classical nature 
of the $XY$ system , we see that, for all $(x,y)$ the following holds
\begin{align*}
    \Tilde{P}_{XY}(x,y)~\Tilde{\rho}_{x,y}^E \leq 
2^{\lambda}~\Tilde{P}_X(x)\cdot P_Y(y)~\Tilde{\rho}^E_x
\end{align*}
where 
\begin{align*}
    \Tilde{\rho}_x^E\coloneqq 
\sum\limits_{y}\Tilde{P}_Y(y|x)\Tilde{\rho}^E_{x,y}
\end{align*}

Coupled with \cref{eq:changinDistributions} this implies that with 
probability at least $1-2\sqrt{\epsilon}$ over the choice of $x$ from 
the distribution $\Tilde{P}_X$ and $y$ from the distribution $P_Y$, the 
following holds
\begin{align}\label{eq:goodCondition}
    \Tilde{\rho}_{x,y}^E \leq 
2^{\lambda}~\frac{1}{1-\sqrt{\epsilon}}~\Tilde{\rho}^E_x
\end{align}

\textbf{The Set of GOOD $x$'s}
Define the function $\mathbf{1}_{x,y}$ as the indicator, which is $1$ 
when \cref{eq:goodCondition} holds. Further, define
\begin{align*}
    \mathbf{1}_x\coloneqq \sum\limits_{y}P_Y(y)\mathbf{1}_{x,y}
\end{align*}
Intuitively, $\mathbf{1}_x$ is the probability that, for a fixed $x$, 
the pairs $(x,y)$ satisfy \cref{eq:goodCondition}, over choice of $y$. 
We know from the discussion in the previous section that
\begin{align*}
    \sum\limits_{x}\Tilde{P}_X(x)\mathbf{1}_x &\geq 1-2\sqrt{\epsilon} 
\end{align*}
Then, another application of Markov's inequality implies that 
\begin{align}\label{eq:goodSet}
    \Pr_{\Tilde{P}_X}\left[ \brak{x~|~\mathbf{1}_x \geq 
1-\epsilon^{1/4}} \right] \geq 1-2\epsilon^{1/4}
\end{align}
We define 
\begin{align*}
    \textsc{nice}_X\coloneqq \brak{x~|~\mathbf{1}_x \geq 1-\epsilon^{1/4}}
\end{align*}

What this implies is that, for any $x\in \textsc{nice}_X$, the 
probability over choice of $y$ that $(x,y)$ satisfies 
\cref{eq:goodCondition} is at least $1-\epsilon^{1/4}$, and that the 
probability that a random $x$ is picked from $\textsc{nice}_X$ is at 
least $1-2\epsilon^{1/4}$ under the distribution $\tilde{P}_X$.

We will however require a few more conditions to define the good set. To 
that end, define
\begin{align*}
    \textsc{nicer}_X\coloneqq \textsc{nice}_X \cap 
\brak{x~|~\norm{P_{Y|x}-P_Y}_1\leq \sqrt{\epsilon}}
\end{align*}
From \cref{eq:diffTildeOriginal} we know that 
\begin{align*}
    \E\limits_{\Tilde{P}_X}\left[ \norm{\Tilde{P}_{Y|X}-P_Y}_1 \right] 
\leq 2\epsilon
\end{align*}
By Markov's inequality we conclude that 
\begin{align*}
    \Pr\limits_{\Tilde{P}_X}\left[ \brak{x~|~\norm{P_{Y|x}-P_Y}_1\leq 
\sqrt{\epsilon}} \right] \geq 1-2\sqrt{\epsilon}
\end{align*}
This implies that 
\begin{align*}
    \Pr_{\Tilde{P}_X}\left[ \textsc{nicer}_X\right] \geq 1-4\epsilon^{1/4}
\end{align*}

Since $\norm{\Tilde{P}_X-P_X}_1\leq \epsilon$, this implies that 
\begin{align*}
    \Pr\limits_{P_X}[\textsc{nicer}_X] \geq 1-5\epsilon^{1/4}
\end{align*}

Next, consider the state 
\begin{align*}
    \rho'\coloneqq \sum\limits_{x,y}P_X(x)P_Y(y) x^X\otimes y^Y\otimes 
\Tilde{\rho}^E_{x,y}
\end{align*}
Then,
\begin{align*}
    \norm{\rho'-\rho}_1 &\leq 
\norm{\rho'-\Tilde{\rho}}_1+\norm{\Tilde{\rho}-\rho}_1 \\
    &= \norm{\Tilde{P}_{XY}-P_X\cdot P_Y}_1 + \norm{\Tilde{\rho}-\rho}_1 
\\ &\leq 3\epsilon \\
\end{align*}
Since
\begin{align*}
    \norm{\rho'-\rho}_1 
=\E\limits_{P_X}\left[\norm{\sum\limits_{y}P_Y(y) y^Y\otimes 
\Tilde{\rho}_{x,y}^E-\sum\limits_{y}P_Y(y) y^Y\otimes \rho_{x,y}^E}_1 
\right]
\end{align*}
this implies that 
\begin{align*}
    \Pr\limits_{P_X}\left[ \norm{\sum\limits_{y}P_Y(y) y^Y\otimes 
\Tilde{\rho}_{x,y}^E-\sum\limits_{y}P_Y(y) y^Y\otimes \rho_{x,y}^E}_1 
\geq \sqrt{\epsilon} \right] \leq 3\sqrt{\epsilon}
\end{align*}

Call the event inside the last probability expression $B_X$. 
Finally, we define 
\begin{align*}
    \textsc{good}_X\coloneqq \textsc{nicer}_X \cap B_X
\end{align*}
This implies that,
\begin{align*}
    \Pr\limits_{P_X}[\textsc{good}_X] \geq 1-10\epsilon^{1/4}
\end{align*}

\textbf{The Covering Lemma}
Let us fix an an $x\in \textsc{good}_X$. Recall that, this implies 
\begin{align*}
    &\norm{\Tilde{P}_{Y|x}-P_Y}_1\leq \sqrt{\epsilon} \\
    \textup{or } 
&\E\limits_{P_Y}\left\lvert\frac{\Tilde{P}_{Y|x}(Y)}{P_Y(Y)}-1\right\rvert 
\leq \sqrt{\epsilon}
\end{align*}
Then, by Markov's inequality,
\begin{align*}
    \Pr\limits_{P_Y}\left[ 
\left\lvert\frac{\Tilde{P}_{Y|x}(Y)}{P_Y(Y)}-1\right\rvert \geq 
\epsilon^{1/4}\right] \leq \epsilon^{1/4}
\end{align*}

Define the set
\begin{align*}
\textsc{good}_{Y|x}\coloneqq \brak{y~|~(x,y) \text{ s.t. } 
\textup{\cref{eq:goodCondition}}, 
\left\lvert\frac{\Tilde{P}_{Y|x}(Y)}{P_Y(Y)}-1\right\rvert \leq 
\epsilon^{1/4}}
\end{align*}

Then, under the distribution $P_Y$, 
\begin{align*}
    \Pr_{P_Y}[\textsc{good}_{Y|x}] \geq 1-2\epsilon^{1/4}
\end{align*}

Define the subdistribution $\bar{P}_Y$ as 
\begin{align*}
    \bar{P}_{Y|x} &= \Tilde{P}_{Y|x} ~~~~~ y\in \textsc{good}_{Y|x} \\
    &= 0 ~~~~~ \text{ otherwise }
\end{align*}

Then it holds that 
\begin{align*}
   \bar{\rho}^{YE}_x\coloneqq 
\sum\limits_{y}\bar{P}_{Y|x}(y|x)~y^Y\otimes \Tilde{\rho}^E_{x,y} &\leq 
2^{\lambda}~\frac{1}{1-\sqrt{\epsilon}}~\big(\sum\limits_{y\in 
\textsc{good}_{Y|x}}\Tilde{P}_{Y|x}(y|x)~y^Y\big)\otimes 
\Tilde{\rho}^{E}_x \\
   &\leq 2^{\lambda}~\frac{1}{1-\sqrt{\epsilon}}~\big(\sum\limits_{y\in 
\textsc{good}_{Y|x}}\Tilde{P}_{Y|x}(y|x)~y^Y+\sum\limits_{y\in \notin 
\textsc{good}_{Y|x}}P_Y(y) y^Y\big)\otimes \Tilde{\rho}^{E}_x \\
   &\leq 
2^{\lambda}~\frac{1}{1-\sqrt{\epsilon}}~
\big((1+\epsilon^{1/4})\sum\limits_{y\in 
\textsc{good}_{Y|x}}P_{Y}(y)~y^Y+\sum\limits_{y\in \notin 
\textsc{good}_{Y|x}}P_Y(y) y^Y\big)\otimes \Tilde{\rho}^{E}_x \\
   &\leq 
2^{\lambda}\frac{1+\epsilon^{1/4}}
{1-\sqrt{\epsilon}}\big(\sum\limits_{y}P_Y(y) 
y^Y\big)\otimes \Tilde{\rho}^E \\
   &= 
2^{\lambda}~\frac{1+\epsilon^{1/4}}{1-\sqrt{\epsilon}}~\rho^Y\otimes 
\Tilde{\rho}^{E}_x
\end{align*}
Also note that
\begin{align*}
    \bar{\rho}_x^{E} &= 
\sum\limits_{y}\bar{P}_{Y|x}(y|x)\Tilde{\rho}^E_{x,y} \\
    &\leq \sum\limits_{y}\Tilde{P}_{Y|x}(y|x)\Tilde{\rho}^E_{x,y} \\
    &= \Tilde{\rho}^E_x
\end{align*}
Using the above fact in the proof of convex split lemma we get that 
\begin{align*}
\norm{\frac{1}{K}\sum\limits_{i=1}^{K}
\bar{\rho}_x^{Y_iE}\bigotimes\limits_{j\neq 
i}\rho^{Y_j}-\Tilde{\rho}^E_x\bigotimes
\limits_{i=1}^{K}\rho^{Y_i}}_1 
\leq 
\sqrt{2^{\lambda}\frac{1+\epsilon^{1/4}}{1-\sqrt{\epsilon}}\frac{1}{K}}
\end{align*}

Next we bound the distance between $\rho^{YE}_x$ and 
$\bar{\rho}^{YE}_x$. To do this, consider the following triangle 
inequality
\begin{align*}
    \norm{\rho-\bar{\rho}}_1 &\leq 
\norm{\bar{\rho}-\sum\limits_{y}\Tilde{P}_{Y|x}(y|x)~y^Y\otimes 
\Tilde{\rho}^E_{x,y}}_1 \\
    &+ \norm{\sum\limits_{y}\Tilde{P}_{Y|x}(y|x)~y^Y\otimes 
\Tilde{\rho}^E_{x,y}-\sum\limits_{y}P_{Y}(y)~y^Y\otimes 
\Tilde{\rho}^E_{x,y}}_1 \\
    &+\norm{\sum\limits_{y}P_{Y}(y)~y^Y\otimes 
\Tilde{\rho}^E_{x,y}-\sum\limits_{y}P_{Y}(y)~y^Y\otimes \rho^E_{x,y}}_1 
\\
    &= \sum\limits_{y:\bar{P}_{Y|x}(y|x)=0}\Tilde{P}_{Y|x}(y) + 
\norm{\Tilde{P}_{Y|x}-P_Y}_1+
\sum\limits_{y}P_Y(y)\norm{\Tilde{\rho}_{x,y}^E-\rho_{x,y}^E}_1
\end{align*}

To bound the first term, observe that the summation is precisely over 
those $y$'s which do not belong to $\textsc{good}_{Y|x}$. We already 
know that under the distribution $P_Y$ this set has probability at most 
$2\epsilon^{1/4}$. Thus by the definition of $\textsc{good}_X$ the first 
term is at most $\sqrt{\epsilon}+2\epsilon^{1/4}\leq 3\epsilon^{1/4}$.

The second and third terms can be bounded similarly directly from the 
definition of $\textsc{good}_X$ by $\sqrt{\epsilon}$ each. Thus,
\begin{align*}
    \norm{\rho^{YE}_x-\bar{\rho}^{YE}_x}_1 \leq 5\epsilon^{1/4}
\end{align*}

We will require one more triangle inequality to replace 
$\Tilde{\rho}^E_x$ with $\rho^E_x$ :
\begin{align*}
    \norm{\Tilde{\rho}^E_x-\rho^E_x}_1 &= 
\norm{\sum\limits_{y}\Tilde{P}_{Y_x}(y|x)\Tilde{\rho}^E_{x,y}-
\sum\limits_{y}P_Y(y)\rho^E_{x,y}}_1 \\
    &\leq \norm{\sum\limits_{y}\Tilde{P}_{Y_x}(y|x)\Tilde{\rho}^E_{x,y}-
\sum\limits_{y}P_Y(y)\Tilde{\rho}^E_{x,y}}_1 \\
    &+ \norm{\sum\limits_{y}P_Y(y)\Tilde{\rho}^E_{x,y}-
\sum\limits_{y}P_Y(y)\rho^E_{x,y}}_1 
\\
    &\leq 2\sqrt{\epsilon}
\end{align*}

Collating all these arguments together and using the standard trick to 
get a covering lemma from the convex split lemma, we see that the 
following holds
\begin{align*}
    \E\limits_{y_1,y_2,\ldots,y_K\sim 
P_{Y}}\norm{\frac{1}{K}\sum\limits_{i}^{K}\rho_{x,y_i}^{E}-\rho^E_x}_1 
\leq 
8\epsilon^{1/4}+\sqrt{2^{\lambda}
\frac{1+\epsilon^{1/4}}{1-\sqrt{\epsilon}}\frac{1}{K}}
\end{align*}

\textbf{The Successive Cancellation Step}

Define $\epsilon_0\coloneqq 10\epsilon^{1/4}$. Suppose we sample $K'$ 
times independently from the distribution $P_X$. Then, by Markov's 
inequality,

\begin{align*}
    \Pr\left[\sum\limits_{i}^{K'}I_{X_i\notin \textsc{good}_X}\geq 
\sqrt{\epsilon_0}\cdot K' \right] \leq \sqrt{\epsilon_0}
\end{align*}

Suppose $x_1x_2\ldots x_{K'}$ is a sequence which has at most 
$\sqrt{\epsilon_0}K'$ samples from $\textsc{good}^c_X$. Then, for this 
fixed sequence $x^{K'}$, the following holds
\begin{align*}
    \E\limits_{y_1,y_2,\ldots,y_K\sim P_{Y}}\norm{\frac{1}{K\cdot 
K'}\sum\limits_{i}^{K'}\sum\limits_{j}^{K}\rho^E_{x_i, y_j}-\rho^E}_1 
&\leq \E\limits_{y_1,y_2,\ldots,y_K\sim P_{Y}} \left[ 
\norm{\frac{1}{K\cdot 
K'}\sum\limits_{i}^{K'}\sum\limits_{j}^{K}\rho^E_{x_i, 
y_j}-\frac{1}{K'}\sum\limits_{i}^{K'}\rho_{x_i}^E}_1 \right] + 
\norm{\frac{1}{K'}\sum\limits_{i}^{K'}\rho_{x_i}^E-\rho^E}_1 \\
    &\leq 
\frac{1}{K'}\sum\limits_{i}^{K'}\E\limits_{y_1,y_2,\ldots,y_K\sim 
P_{Y}}\left[ \norm{\frac{1}{K}\sum\limits_{j}^{K}\rho^E_{x_i, 
y_j}-\rho_{x_i}^E}_1 \right]+ 
\norm{\frac{1}{K'}\sum\limits_{i}^{K'}\rho_{x_i}^E-\rho^E}_1 \\
    & \leq \frac{1}{K'}\Big( 
(1-\sqrt{\epsilon_0})K'\cdot\left(8\epsilon^{1/4}+
\sqrt{2^{\lambda}\frac{1+\epsilon^{1/4}}
{1-\sqrt{\epsilon}}\frac{1}{K}}\right)+\sqrt{\epsilon_0}K'\cdot 
2 \Big) \\
    &+ \norm{\frac{1}{K'}\sum\limits_{i}^{K'}\rho_{x_i}^E-\rho^E}_1
\end{align*}

We will now set the values of $K$ and $K'$. We set $K$ and $K'$ such that 
\begin{align*}
    &2^{\lambda}\frac{1+\epsilon^{1/4}}{1-\sqrt{\epsilon}}\frac{1}{K} 
\leq \epsilon^{1/4} \\
    \intertext{and }
    &\E\limits_{x_1,x_2,\ldots,x_{K'}\sim 
P_{X}}\norm{\frac{1}{K'}\sum\limits_{i}^{K'}\rho_{x_i}^E-\rho^E}_1 \leq 
\epsilon^{1/4}
\end{align*}

The second inequality can be set by using the smoothed version of the 
convex split lemma.
Then, 
\begin{align*}
    \E\limits_{\substack{x_1,x_2,\ldots,x_{K'}\sim P_{X} \\ 
y_1,y_2,\ldots,y_K\sim P_{Y}}}\norm{\frac{1}{K\cdot 
K'}\sum\limits_{i}^{K'}\sum\limits_{j}^{K}\rho^E_{x_i, y_j}-\rho^E}_1 
&\leq \Big ((1-\sqrt{\epsilon_0})\cdot 
9\epsilon^{1/4}+2\sqrt{\epsilon_0}\Big)\cdot 
(1-\sqrt{\epsilon_0})+2\sqrt{\epsilon_0}+\epsilon^{1/4} \\
    &\leq 10\epsilon^{1/8}
\end{align*}
\end{proof}
\vspace{2mm}
\begin{proof}[Proof of \cref{lem:chainRule}]
Suppose that $\ket{\varphi}^{RABC}$ is a purification of 
$\varphi^{RAB}$, where $C$ is the purifying register. Consider the 
following task : Let Alice possess the systems $ABC$ and $R$ be the 
reference. Alice wants to send the systems $AB$ to Bob. This is known as 
quantum \emph{state splitting}. We will achieve this task in two steps. 
In Step $1$, Alice will send the system $A$ to Bob while treating $BC$ 
as the purifying registers. This will require 
$\frac{1}{2}I_{\max}^{\epsilon}(R:A)_{\varphi}+\log \frac{1}{\epsilon}$ 
bits of quantum communication. In Step $2$, Alice will send the system 
$B$ to Bob, while while treating the system $C$ as the purifying 
register. This task will require 
$\frac{1}{2}I_{\max}^{\epsilon}(RA:B)+\log\frac{1}{\epsilon}$ bits of 
quantum communication. At the end of the protocol Alice will have 
successfully sent Bob the systems $AB$, with some $O(\epsilon)$ error. 
We already know from \cite{bertaReverseShannon} that any one-way 
entanglement assisted protocol that achieves this task with $\epsilon$ 
error requires at least $\frac{1}{2}I_{\max}^{\epsilon}(R:AB)_{\varphi}$ 
number of qubits. Collating these arguments together gives us the upper 
bound.

To achieve Step $1$ and Step $2$ above, we will use the smoothed convex 
split lemma, specifically the protocol in Theorem 1 of 
\cite{convexSplit}.

\textbf{Step 1}

Let $\varphi'^{RA}$ be the optimiser for the expression 
$\Tilde{I}_{\max}^{\epsilon}(R:A)_{\varphi}$. Then the smoothed convex 
split lemma, along with two triangle inequalities shows us that

\begin{align*}
  P(\frac{1}{n}\sum\limits_{i=1}^{n}\varphi^{RA_i}\bigotimes\limits_{j\neq 
i}\varphi^{A_j}, \varphi^{R}\bigotimes\limits_{i=1}^{n}\varphi^{A_i}) 
\leq 3\epsilon
\end{align*}
where $n> \Tilde{I}_{\max}^{\epsilon}(R:A)_{\varphi}+
	2\log \frac{1}{\epsilon}$. 

Armed with this relation, we can directly use the protocol in Theorem 1 
to send the system $A$ to Bob and obtain a pure state 
$\ket{\varphi"}^{RABC}$ such that
\begin{enumerate}
    \item $P(\varphi"^{RABC},\varphi^{RABC})\leq 3\epsilon$
    \item The system $A$ is now with Bob.
\end{enumerate}

\textbf{Step 2}
For the next part of the protocol, we recall that, whenever $m> 
\Tilde{I}_{\max}^{\epsilon}(RA:B)_{\varphi}+2\log\frac{1}{\epsilon}$
\begin{align*}
    P(\frac{1}{m}\sum\limits_{i=1}^{m}\varphi^{RAB_i}\bigotimes
\limits_{j\neq 
i}\varphi^{B_j}, \varphi^{RA}\bigotimes\limits_{i=1}^{m}\varphi^{B_i}) 
\leq 3\epsilon
\end{align*}

However, since we global state shared by Alice, Bob and Referee is 
$\varphi"$, we need a further triangle inequality to show that
\begin{align*}
   P(\frac{1}{m}\sum\limits_{i=1}^{m}\varphi"^{RAB_i}
	\bigotimes\limits_{j\neq 
i}\varphi^{B_j}, \varphi"^{RA}\bigotimes\limits_{i=1}^{m}\varphi^{B_i}) 
\leq 9\epsilon
\end{align*}

Repeating the protocol in Theorem 1 to send the system $B$ to Bob, while 
Bob possesses the system $A$, we get a pure state 
$\Tilde{\varphi}^{RABC}$ such that
\begin{align*}
     &P(\Tilde{\varphi}^{RABC},\varphi"^{RABC})\leq 9\epsilon \\
     \implies & P(\Tilde{\varphi}^{RABC},\varphi^{RABC}) \leq 12\epsilon
\end{align*}
Along with the lower bound, this implies that 
\begin{align*}
    \Tilde{I}_{\max}^{\epsilon}(R:A)_{\varphi}+
\Tilde{I}_{\max}^{\epsilon}(RA:B)_{\varphi}+4\log 
\frac{1}{\epsilon} \geq \Tilde{I}_{\max}^{\epsilon}(R:AB)_{\varphi}
\end{align*}

Finally, we use the bound that, for any state $\rho^{AB}$
\begin{align*}
    \Tilde{I}_{\max}^{\epsilon}(A:B)_{\rho}\leq 
I_{\max}^{\epsilon-\gamma}(A:B)_{\rho}+\log\frac{3}{\gamma^2}
\end{align*}
to get the desired chain rule.

\end{proof}
\vspace{2mm}
\begin{proof}[Proof of \cref{lem:3senderMac}]

The proof of this lemma is a simple Fourier-Motzkin elimination, with 
the extra condition that

\begin{align*}
    R_1=R_{10}+R_{11} \end{align*} and hence we omit it. \end{proof} 
\vspace{2mm} \begin{proof}[Proof of \cref{thm:mainPrivateRegion}] 
Fix a 
$\theta\in [0,1]$ and fix the rate tuple for Alice\sub{1}, Bob and 
Alice\sub{2}, $(R_{10},R_2, R_{10})$ such that it belongs to the region 
given by \cref{eq:rateRegion}.
By \cref{lem:3senderMac}, this ensures that the rate pair 
$(R_{10}+R_{11}, R_2)$ is achievable for classical message transmission 
across the cq-mac, with error at most $\epsilon^{1/8}$. Recall that we 
use the definition $R_1=R_{10}+R_{11}$.

Again, fix $K_1, K_2$ and $K_3$ i.e. the block sizes over which 
Alice\sub{1}, Bob and Alice\sub{2} randomise, as in 
\cref{thm:mainSuccCanccCovering}. Define

\begin{align*}
    &R^{private}_{\textsc{Alice}}\coloneqq R_1-\log K_1-\log K_2 \\
    &R^{private}_{\textsc{Alice}}\coloneqq R_2-\log K_3
\end{align*}

The code construction is as follows:
\begin{enumerate}

    \item For Alice\sub{1} choose symbols $x(1),x(2),\ldots,x(R_{10})$ 
iid from $P_U^{\theta}$.

    \item For Bob choose symbols $y(1),y(2),\ldots,y(R_2)$ iid from 
$P_Y$.

    \item For Alice\sub{2} choose symbols $z(1),z(2),\ldots,z(R_{11})$ 
iid from $P_V^{\theta}$.

    \item Divide Alice\sub{1}'s codebook into blocks, each of size 
$K_1$. Do the same for Bob and Alice\sub{2}, with block sizes $K_2$ and 
$K_3$ respectively.

    \item Alice\sub{1} maps her message set $[M_1]$ to codebook such 
that each message $m_1\in [M_1]$ corresponds to some block. Bob and 
Alice\sub{2} do the same, for their message sets $[N]$ and $[M_2]$.

    \item To send the message $m_1$, Alice\sub{1} goes into the block 
corresponding to that message. Suppose that block contains the symbols 
$(x_{m_1}(1),x_{m_1}(2),\ldots, x_{m_1}(K_1))$, Alice\sub{1} picks a 
symbol uniformly at random and transmits it,

    \item Bob and Alice\sub{2} do the same for their corresponding 
messages $n$ and $m_2$.

\end{enumerate}

Decodability is guaranteed by the code specified rates and 
\cite{sen2020unions}. Secrecy is guaranteed by the values of $K_1, K_2, 
K_3$ and \cref{thm:mainSuccCanccCovering}. This argument implies that 
for any rate tuple $(R_1,R_2)$ that lies in the region given by 
\cref{lem:3senderMac}, and for any $(\log K, \log K')$ such that

\begin{align*}
    \log K \geq \log K_1 + \log K_3 \\
    \log K' \geq \log K_2
\end{align*}

the rate pair $(R_1-\log K, R_2-\log K')$ is achievable for private 
transmission across the cq mac, assuming both coordinates are 
non-negative. This is precisely the definition of the set 
$\left(\mathcal{S}_{\theta}-\mathcal{T}_{\theta}\right)^+$. Repeating 
this procedure for all $\theta\in [0,1]$ and then taking a union bound 
over all the regions concludes the proof.

\end{proof}
\begin{proof}[Proof of \cref{thm:asymIID}]

The proof is easy and we only provide a brief sketch. First note that 
using \cref{lem:asymIID}, it is easy to see (using the chain rule for 
the mutual information and the data processing inequality) that for 
\emph{every} $\theta\in [0,1]$, the region $\mathcal{S}_{\theta}$ is 
equivalent to the region

\begin{align*}
    &R_1 < I(X:YC) \\
    &R_2 < I(Y:XC) \\
    &R_1+R_2 < I(XY:C)
\end{align*}

Call this region $\mathcal{S}$. Along with \cref{thm:mainPrivateRegion}, 
this implies that the private capacity region is given by

\begin{align*}
    \left(\mathcal{S}-\bigcup\limits_{\theta\in [0,1]} 
\mathcal{T}_{\theta}\right)^{+}
\end{align*}

Using the continuity of the mutual information with respect to 
$\theta\in [0,1]$, and again via the chain rule for the mutual 
information, we see that the region $\bigcup\limits_{\theta\in [0,1]} 
\mathcal{T}_{\theta}$ is equivalent to
\begin{align*}
    &\log K \geq I(X:E) \\
    & \log K' \geq I(Y:E) \\
    & \log K +\log K' \geq I(XY:E)
\end{align*}

Taking the difference of these two regions gives us the desired region 
for private transmission. This concludes the proof.
\end{proof}

\bibliography{ref3}
\bibliographystyle{alpha}

\end{document}